\documentclass[prl,superscriptaddress,twocolumn]{revtex4-1}
\pdfoutput=1
\usepackage[colorlinks=true,urlcolor=blue]{hyperref}
\usepackage{amsfonts}
\usepackage{graphicx}
\usepackage{mathrsfs}
\usepackage{amsmath}
\usepackage{upgreek}
\usepackage{xcolor}
\usepackage{bm}
\usepackage{titlesec}
 
\usepackage{color}
\definecolor{red}{rgb}{0.75,0,0}
\definecolor{blue}{rgb}{0,0,0.75}
\definecolor{green}{rgb}{0,0.5,0}

\titlespacing*{\section}
{0pt}{0.5cm}{0.25cm}
\titlespacing*{\subsection}
{0pt}{0.5cm}{0.25cm}

\begin{document}

\title{Mono- to Multilayer Transition in Growing Bacterial Colonies}

\author{Zhihong You}
\affiliation{Instituut-Lorentz, Universiteit Leiden, P.O. Box 9506, 2300 RA Leiden, Netherlands}
\author{Daniel J. G. Pearce}
\affiliation{Instituut-Lorentz, Universiteit Leiden, P.O. Box 9506, 2300 RA Leiden, Netherlands}
\author{Anupam Sengupta}
\affiliation{Physics and Materials Science Research Unit, University of Luxembourg, 162 A, Avenue de la Faïencerie, Luxembourg City, L-1511 Grand Duchy of Luxembourg}
\author{Luca Giomi}
\email{giomi@lorentz.leidenuniv.nl}
\affiliation{Instituut-Lorentz, Universiteit Leiden, P.O. Box 9506, 2300 RA Leiden, Netherlands}

\begin{abstract}
\vspace*{0.5cm}
The transition from monolayers to multilayered structures in bacterial colonies is a fundamental step in biofilm development. Observed across different morphotypes and species, this transition is triggered within freely growing bacterial microcolonies comprising a few hundred cells. Using a combination of numerical simulations and analytical modeling, here we demonstrate that this transition originates from the competition between growth-induced in-plane active stresses and vertical restoring forces, due to the cell-substrate interactions. Using a simple chainlike colony of laterally confined cells, we show that the transition sets when individual cells become unstable to rotations, thus it is localized and mechanically deterministic. Asynchronous cell division renders the process stochastic, so that all the critical parameters that control the onset of the transition are continuously distributed random variables. Here we demonstrate that the occurrence of the first division in the colony can be approximated as a Poisson process in the limit of large cells numbers. This allows us to approximately calculate the probability distribution function of the position and time associated with the first extrusion. The rate of such a Poisson process can be identified as the order parameter of the transition, thus highlighting its mixed deterministic-stochastic nature.
\end{abstract}

\maketitle


Bacteria dividing on a substrate first give rise to an exponentially growing flat monolayer of packed and partially aligned cells, which upon reaching a critical population size, invades the third dimension resulting in a growing colony of multiple bacterial layers \cite{Berne:2018,Su:2012,Grant:2014,Duvernoy:2018,Beroz:2018}. While the viability of bacterial populations over these stages is determined by a complex interplay of biochemical cues, the colony structure and the dynamics of the cell in 2D to 3D colony transitions are mediated by biophysical factors, including the division rate, cell-to-cell and the cell-to-surface interactions within bulk \cite{Fuentes:2013,Grant:2014}, and bounded \cite{Volfson:2008,Boyer:2011,Sheats:2017} or free boundary conditions \cite{You:2018,Arciprete:2018}.

Transitions from mono- to multilayered structures have recently drawn significant attention in the biophysical literature, being a universal step toward the formation of multicellular biofilm structures, as well as a process where mechanical forces are likely to play a leading role. Grant {\em et al.} \cite{Grant:2014}  investigated the mono- to multilayer transition in {\em E. coli} colonies confined between glass and agarose and found that the size of the colony at onset is affected by the substrate stiffness and friction. More recently, Beroz {\em et al}. \cite{Beroz:2018} demonstrated that, in {\em V. cholerae} biofilms, escape to the third dimension is mediated by a {\em verticalization} of the longer cells. Similar mechanisms are also found in confluent monolayers of eukaryotic cells \cite{Marinari:2012,Eisenhoffer:2012,Guillot:2013,Kocgozlu:2016,Saw:2017} and are believed to regulate cell extrusion and apoptosis. 

Despite these works having greatly contributed to shed light on the nature of the transition, some aspects are still debated \cite{Allen:2019}. Is there a well defined critical size, stress, and time at which extrusion is first triggered? Is the mono- to multilayer transition a deterministic process, or does it result from an interplay of deterministic and stochastic effects? To what extent can this process be likened to a nonequlibrium phase transition?

In this Letter we address these questions theoretically, using a combination of numerical and analytical methods. We show that the mono- to multilayer transition in a system of growing rodlike cells results from a competition between the in-plane active stresses, that compress the cells laterally, and the vertical restoring forces, owing to the cell-substrate interactions (e.g. cell-substrate adhesion). As the colony expands the internal stress increases until it is sufficiently large to cause extrusion of the first cells. In the ideal case of a laterally confined chainlike colony of non-growing cells subject to axial compression, the transition is entirely deterministic and the critical stress at which extrusion initiates can be calculated analytically. Asynchronous cell division, however, renders the transition stochastic. In this case, the critical stress is a continuously distributed random variable and the first extrusion does not necessarily occur at the colony center, despite this being the region of maximal stress. Upon modeling cell division as a Poisson process, we can approximately calculate the probability distribution function (PDF) of the position and time associated with the first extrusion. Finally, we show that the rate of the Poisson process is analogous to an order parameter and that, in this respect, the mono- to multilayer instability is likened to a continuous phase transition.

\begin{figure}[t]
\centering
\includegraphics[width=\columnwidth]{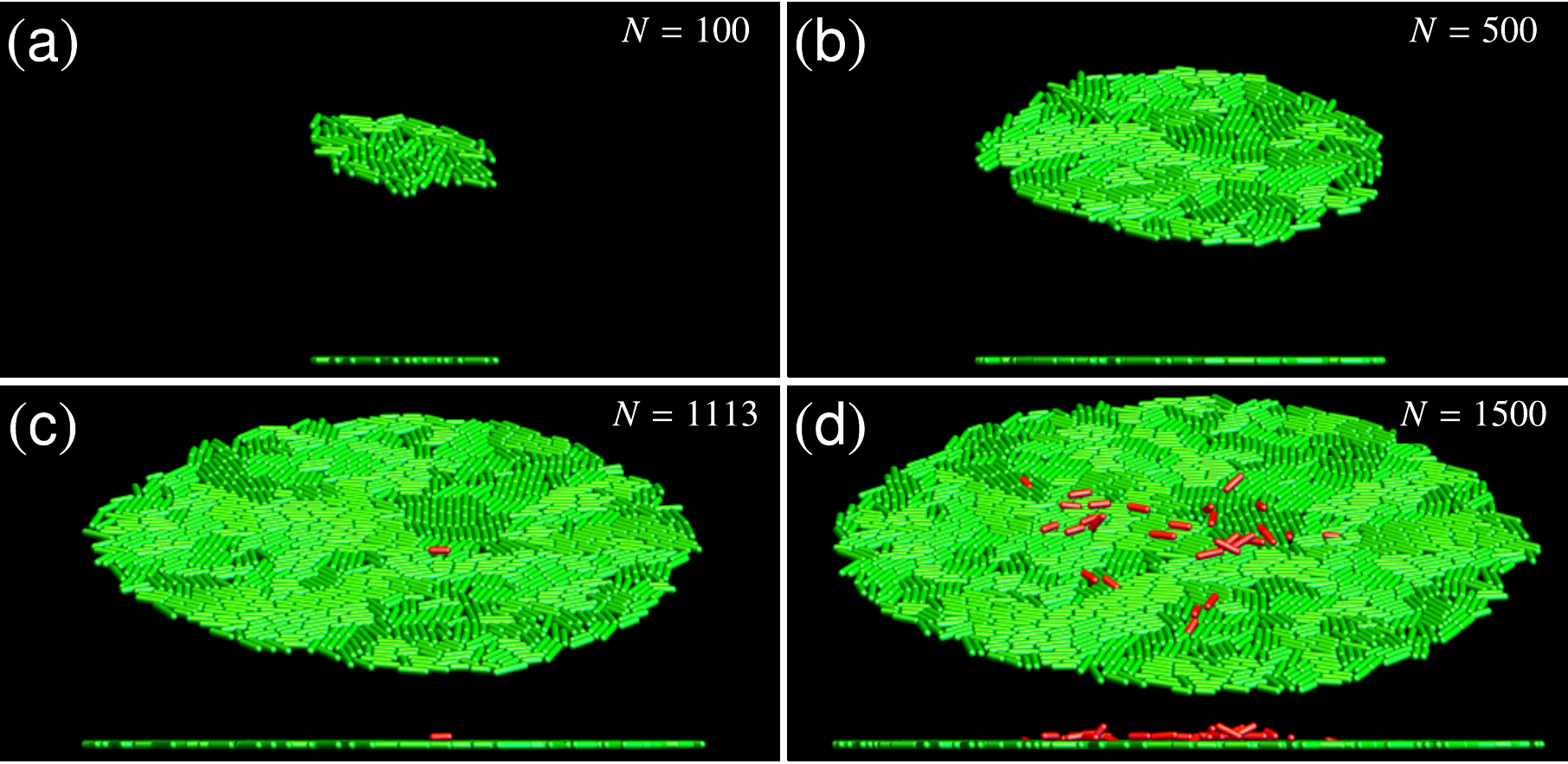}
\caption{\label{fig:snapshots} (a)--(d) Snapshots of a simulated growing colony at different ages to show the mono- to multilayer transition. The lower image in each panel shows the side view. In panels (c) and (d), the extruded cells are highlighted as red.}
\end{figure}

We employ a toy model of duplicating bacteria \cite{Farrell:2013,You:2018,Ghosh:2015}, where cells are represented as spherocylinders with a fixed diameter $d_{0}$ and a time-dependent length $l$ (excluding the caps on both ends), growing in three-dimensional space and subject to Brownian motion \cite{SI}. Whereas cells in bacterial colonies are potentially subject to a large variety of mechanical and biochemical stimuli, here we focus on three types of forces: the repulsive forces associated with cell-cell and cell-substrate steric interactions and a vertical restoring force, representing either the attractive force due to adhesion of the cells with the extracellular matrix (ECM) \cite{Beroz:2018}, or a mechanical compression from above \cite{Grant:2014,Duvernoy:2018}. All forces are treated as Hookean. For simplicity, we assume cell-cell and cell-substrate repulsive interactions to be characterized by the same elastic constant $k=10$ ${\rm MPa}\,\upmu{\rm m}$, while we set the elastic constant associated with adhesion to be $k_{a}l$, to mimic the dependence of the restoring forces on the contact area. The length $l_{i}$ of the $i$th cell increases in time with rate $g_{i}$ and, after having reached the value $l_{d}$, the cell divides into two identical daughter cells. To avoid synchronous divisions, the growth rate of each cell is randomly chosen in the interval $g/2\le g_{i} \le 3g/2$, with $g$ the average growth rate. Our results remain unchanged in case of Gaussianly distributed growth rate \cite{SI}. In-depth studies and more realistic models of bacterial growth can be found, e.g., in Refs. \cite{Wang:2010,Amir:2014,TaheriAraghi:2015}. We stress that our model does not aim to accurately reproduce the traits of a specific bacterial family, but rather to abstract the essential features that all bacterial species undergoing the mono- to multilayer transition have in common. Figure ~\ref{fig:snapshots} shows typical configurations of our {\em in silico} colonies at different time points. Consistent with the experimental evidence \cite{Su:2012,Grant:2014}, the colony initially expands as a perfect monolayer [Figs. ~\ref{fig:snapshots}(a) and ~\ref{fig:snapshots}(b)] and, once it is sufficiently large, some cells are extruded and originate a second layer [Figs. ~\ref{fig:snapshots}(c) and ~\ref{fig:snapshots}(d)]. See Ref.~\cite{SI} for time-lapse animations showing the growth dynamics of the colonies.

\begin{figure}[t]
\centering
\includegraphics[width=\columnwidth]{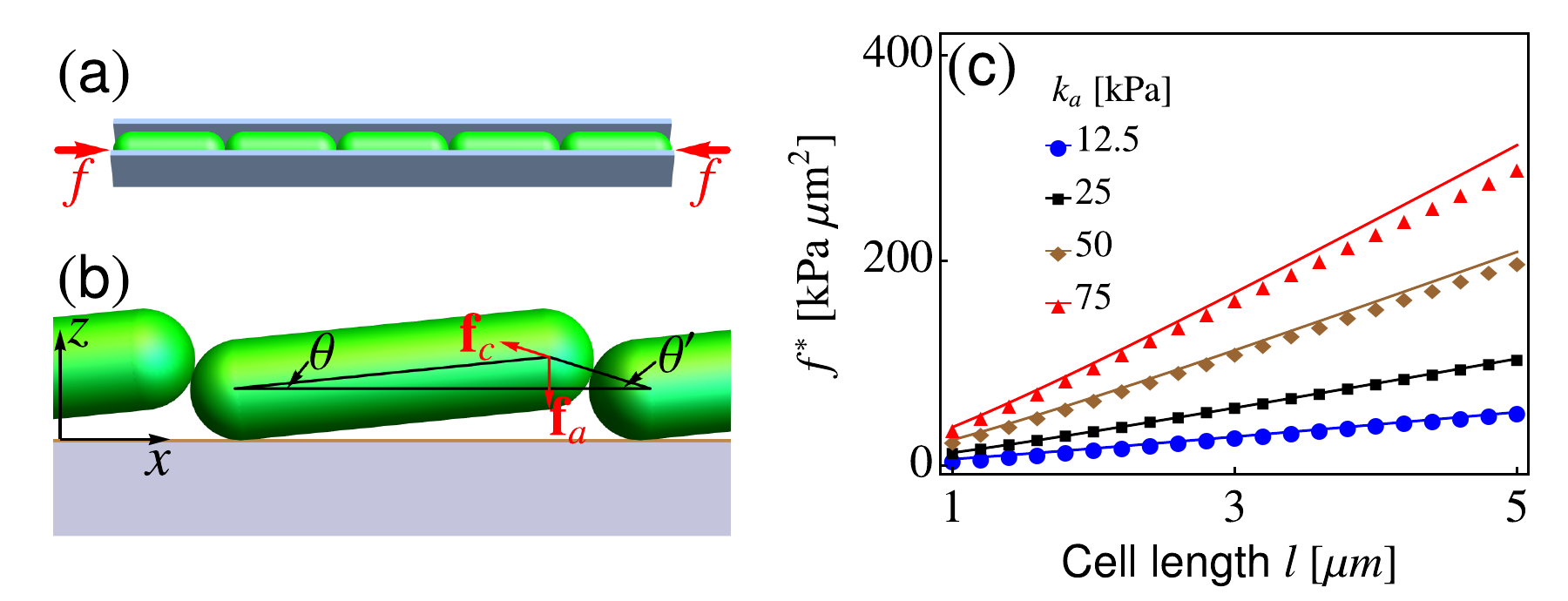}
\caption{\label{fig:mechanics} 
(a) Schematic diagram of the chainlike colony. (b) Schematics of torque balance about the lower end of the cell. (c) Critical force as a function of the cell length, for various $k_{a}$ values. The dots and lines represent, respectively, the simulation and analytical results as in Eq. \eqref{eq:fc}.}
\end{figure}

As a starting point, we look at a simplified chainlike colony, consisting of a row of cells confined in a channel [Fig. \ref{fig:mechanics}(a)]. The cells have identical length $l$ and do not grow, but are compressed by a pair of forces $f$ applied at the two ends of the channel. As in the case of planar colonies (Fig. \ref{fig:snapshots}), cells remain attached to the substrate for small compressive forces and are extruded to the second layer for large $f$ values. In this case, the transition is entirely deterministic and there exists a well-defined critical force, $f^{*}$, at which the monolayer becomes unstable. This can be calculated analytically upon balancing the torques associated with cell-cell and cell-substrate interactions, about the lower end of the cell axis. Calling $\bm{p}=(\cos\theta,0,\sin\theta)$ the orientation of the first extruded cell and $\bm{f}_{c}=f_{c}(-\cos\theta',0,\sin\theta')$, with $f_{c}=f/\cos\theta'$, the contact force exerted by the nearby cell [Fig. \ref{fig:mechanics}(b)], the lifting torque can be calculated in the form $\tau_{c}=l(p_{x}f_{z}-p_{z}f_{x})=lf\cos\theta(\tan\theta+\tan\theta')$. Analogously, the restoring torque resulting from the adhesive force is $\tau_{a}=k_{a}l^{3}\sin \theta\cos \theta$. In a perfectly horizontal monolayer, $\theta=\theta_{0}=0$ and both torques vanish. In order for such a configuration to be stable against slight orientational fluctuations of magnitude $\delta \theta \ll 1$, $\tau_{c}(\theta_{0}+\delta\theta)<\tau_{a}(\theta_{0}+\delta\theta)$. Upon expanding $\tau_{c}$ and $\tau_{a}$ at the linear order in $\delta\theta$ and approximating $\theta'\approx(l/d_{0})\theta$, one can verify that such a stability condition breaks down when $f>f^{*}$, with
\begin{equation}
\label{eq:fc}
f^{*}=\frac{k_{a}l^{2}}{1+l/d_{0}}\;,
\end{equation}
in excellent agreement with the result of our numerical simulations [Fig. \ref{fig:mechanics}(c)]. The existence of a well-defined critical force resulting from the competition between compression and rotation is vaguely reminiscent of Euler's buckling in elastic rods. However, while buckling is a system-wide instability, the mono- to multilayer transition is determined by torque balance at the length scale of a single cell.


Next we explore the effect of asynchronous cell division. Cells are again confined in the channel and, unlike the previous case, they are not subject to lateral compression, but elongate and divide. To investigate the effect of the key parameters, $k_{a}$, $l_{d}$, and $g$, we perform four sets of $10^{4}$ simulations, starting from a single cell at the equilibrium configuration. In the ``control'' set, we fix $k_{a}=25$ kPa, $l_{d}=4\ \upmu$m, and $g=2\ \upmu$m/h. In each of the remaining three sets we change one of the parameters. 

\begin{figure}[t]
\centering
\includegraphics[width=\columnwidth]{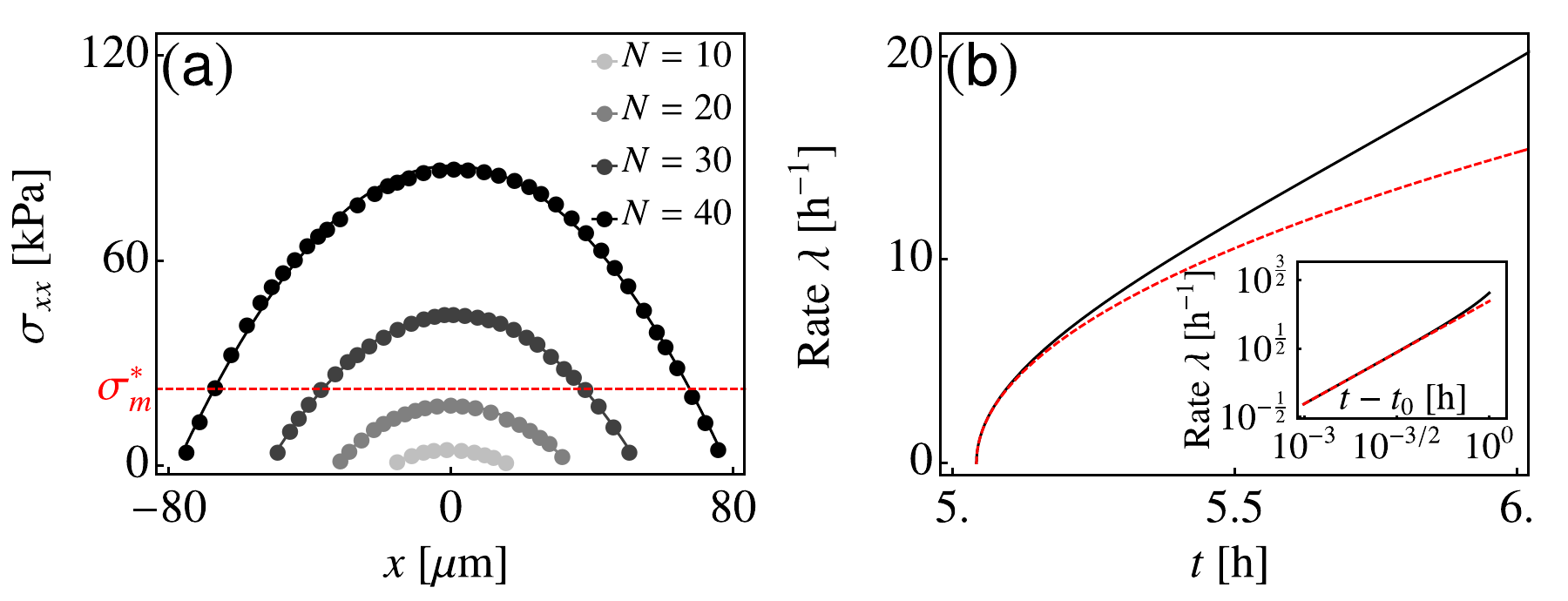}
\caption{\label{fig:stre-lamb} (a) The spatial distributions of stress in a growing chainlike colony at different ages. (b) Rate $\lambda(t)$ of the Poisson process at $\Delta t = 0$. The black and red dashed lines represent, respectively, Eq. \eqref{eq:lambt} and $\lambda(t)\approx k_{\lambda}\sqrt{t-t_{0}}$. The inset shows the same plot in a log-log scale. Both panels are obtained using the ``control'' parameters.}
\end{figure}

\begin{figure*}[t]
\centering
\includegraphics[width=1.0\textwidth]{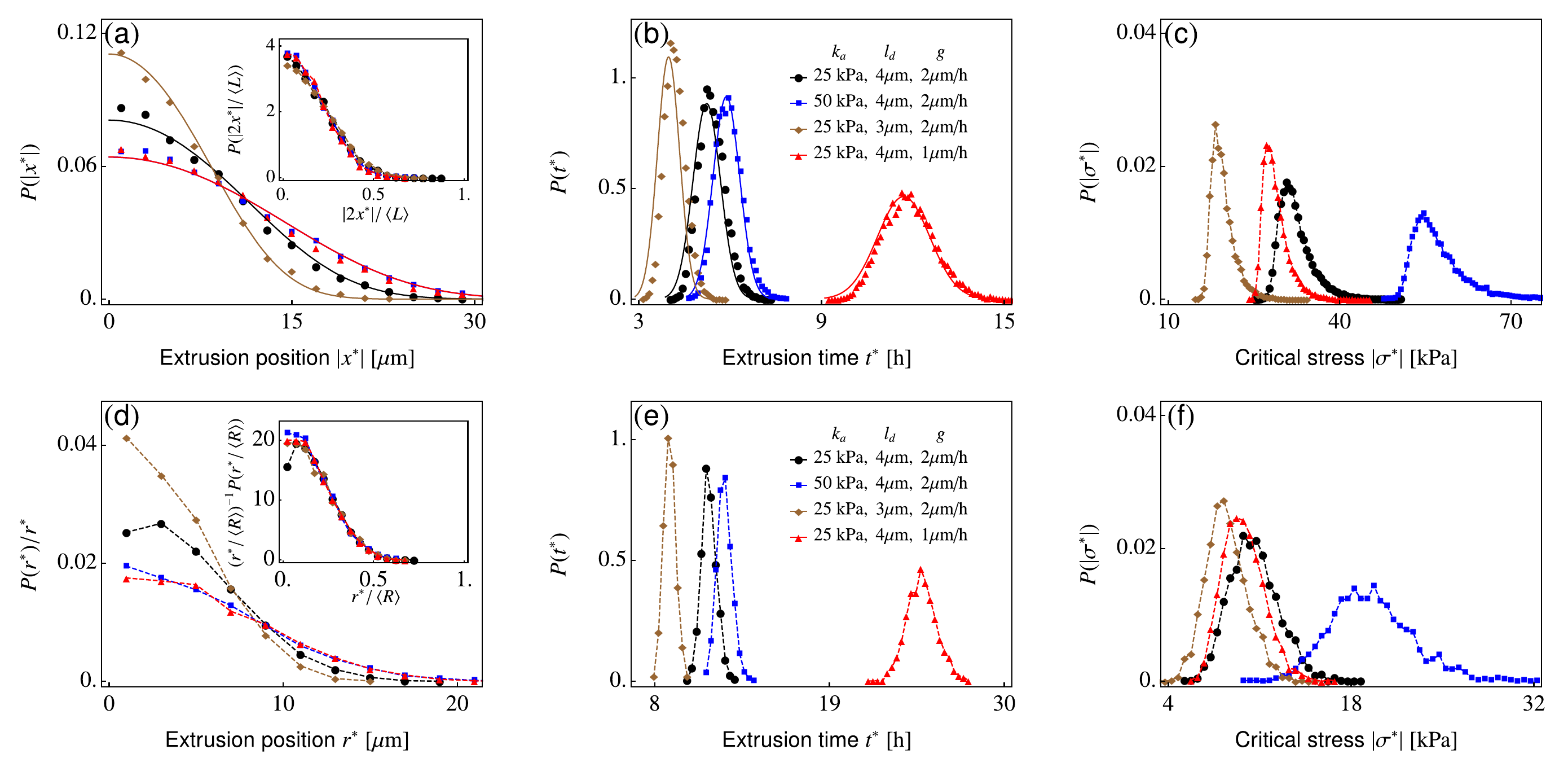}
\caption{\label{fig:statistics}
(a)-(c) Probability densities of (a) the extrusion positions $|x^{*}|$ (inset shows that of the rescaled position $|2x^{*}|/\left\langle L \right\rangle$), (b) the extrusion time $t^{*}$, and (c) the critical stress $\sigma^{*}$, for chainlike colonies of asynchronously dividing cells. (d) Probability density of the extrusion position $r^{*}$ in planar colonies, normalized by $r^{*}$, the distance from the point of extrusion to the centroid of the colony. (e),(f) Same as panels (b),(c), but for planar colonies. In all panels, dots and dashed lines correspond to the simulation results and the solid lines to the analytical predictions. The statistical results are collected at four sets of parameters, whose values are shown in panels (b) and (e). At each set of parameters, there are $10^{4}$ runs for chainlike colonies, and $2\times 10^{3}$ runs for planar colonies.}
\end{figure*}

As the colony expands, the longitudinal stress (calculated via the virial construction \cite{SI}) progressively builds up, while preserving a simple parabolic profile of the form
\begin{equation}
\label{eq:sigmar}
\sigma_{xx}(x) = \sigma_{m}\left[ 1-\left( \frac{2x}{L} \right)^{2} \right]\;,
\end{equation}
where $\sigma_{m}$ and $L$ represent, respectively, the maximum stress and the colony length [Fig. \ref{fig:stre-lamb}(a)]. One can show that $\sigma_{m}=aN^{2}$ and $L=bN$ [Fig. (S3)], where $N$ is the total number of cells and $a$ and $b$ are constants depending on the mechanical properties of the cells and the substrate \cite{SI}. Because the stress is maximal at the center of the colony, one would expect the first extrusion to occur here. Our simulations, however, show a dramatically different behavior. Specifically, the position of the first extruded cell $x^{*}$ follows a broad distribution, whose spread is comparable to the size of the colony itself [Fig. \ref{fig:statistics}(a)]. Remarkably, this distribution depends on the material parameters only through the size $L$ of the colony and all the data collapse onto the same curve upon rescaling $x^{*}$ by the average colony size $\langle L \rangle$ [Fig. \ref{fig:statistics}(a) inset]. Analogously the transition time $t^{*}$ [Fig. \ref{fig:statistics}(b)] and the critical stress $\sigma^{*}$ experienced by cells at the verge of extrusion [Fig. \ref{fig:statistics}(c)], are continuously distributed random variables.

In the following, we demonstrate that, in our model of bacterial colonies, this behavior results from the combined inherent randomness of the division process and the local nature of the instability. According to Eq. \eqref{eq:fc} a cell is unstable to extrusion if subject to a critical force whose magnitude increases with the cell length. In a growing colony, a division event introduces a sudden drop in the cell length and this can, in turn, trigger an extrusion instability, as long as the cell is subject to a stress larger than that required to extrude a cell of minimal length $l_{m}=(l_{d}-d_{0})/2$. We denote such a minimal critical stress $\sigma_{m}^{*}$. As the stress is spatially inhomogeneous and increasing in time, there will be a whole region, symmetric with respect to the center of the colony and whose length increases in time, where the local stress exceeds $\sigma_{m}^{*}$ and cell division can trigger the first extrusion. We call this region the P zone. The probability associated with the first extrusion is then equal to the probability of having a division within the P zone. This can be calculated as follows. 

Let us consider a colony of $n$ cells with growth rate $g$ and assume that, at an arbitrary time, their lengths are independent and uniformly distributed in the interval $l_{m} \le l \le l_{d}$. After a time $t$, the probability that no division has yet occurred equates the probability that none of the cells is initially longer than $l_{d}-gt$:
\begin{equation}\label{eq:pnodivi1}
P(t) = \left(\frac{l_{d}-gt-l_{m}}{l_{d}-l_{m}}\right)^{n} \approx e^{-\lambda(n)t}\;,	
\end{equation}
where $\lambda(n)=ng/(l_{d}-l_{m})$ and the approximation holds for large $n$ values. Equation \eqref{eq:pnodivi1} defines a Poisson process of rate $\lambda(n)$ \cite{Kingman:1992}. If $n$ is time dependent, the process becomes inhomogeneous, but the probability preserves the same structure, with $\lambda(t)=\lambda[n(t)]$ and $P(t)=e^{-\int_{0}^{t}dt'\,\lambda(t')}$. The PDF associated with observing the first division at time $t$ is then
\begin{equation}\label{eq:f}
f(t)=\frac{d}{dt}\,[1-P(t)]=\lambda(t)e^{-\int_{0}^{t}dt'\,\lambda(t')}\;.
\end{equation}
Interestingly, an analogous Poissonian PDF has been also postulated by Allen and Waclaw in a recent review article \cite{Allen:2019}. Equation \eqref{eq:f} supports this conjecture and further provides it with a mechanistic interpretation. To make progress, one needs to calculate the number of cells $n$ within the P zone. This is, on average, $n=L^{*}/(\langle l \rangle+d_{0})$, where $L^{*}$ is the length of the P zone and $\langle l \rangle=(l_{d}+l_{m})/2$ the average cell length. $L^{*}$ can be calculated by solving $\sigma_{xx}(L^{*}/2)=\sigma_{m}^{*}$ [red dashed line in Fig. \ref{fig:stre-lamb}(a)]. This yields: $L^{*} = b \sqrt{N^{2}(t)-N^{2}_{0}}$, while $N_0=\sqrt{\sigma_{m}^{*}/a}$ is the minimal number of cells required for the P zone to exist. From this and Eq. \eqref{eq:pnodivi1}, we can calculate the rate $\lambda(t)$ as:
\begin{equation} \label{eq:lambt}
\lambda(t) = \frac{gb}{(l_{d}-l_{m})(\langle l \rangle+d_{0})}\,\sqrt{N^{2}(t)-N_{0}^{2}} \sim [N(t)-N_0]^{1/2}\;.
\end{equation}
Equation \eqref{eq:lambt} highlights the role of $\lambda$ as order parameter for the mono- to multilayer transition. For $N(t)<N_{0}$, $\lambda$ is imaginary and the probability of observing an extrusion vanishes identically. On the other hand, for $N(t)>N_{0}$, $\lambda$ is real and the probability of observing an extrusion increases in time. The transition is continuous in this case, but other scenarios are likely possible.  

To make the time dependence explicit in Eq. \eqref{eq:lambt}, we need to calculate $N(t)$. Evidently, the average number of cells in the colony grows exponentially in time. Because cells have random growth rates, the time $t$ taken for the colony to attain a given population size $N(t)$, is a random variable of the form $t=\bar{t}+\Delta t$ (Fig. S4 in Ref. \cite{SI}). Numerically, we find that $\Delta t$ approximately follows a Gaussian distribution, $\mathcal{N}(0,\delta_{\Delta t}^{2})$, having zero mean and whose variance, $\delta_{\Delta t}^{2}$, depends on $l_{d}$ and $g$ \cite{SI}. Taking $N(t)=\exp[\omega(t-\Delta t)]$, with $\omega$ a constant, and using Eq. \eqref{eq:lambt}, yields an expression for $\lambda(t)$, hence for $f(t)=f(t|\Delta t)$. A plot of the rate $\lambda(t)$ is shown in Fig. \ref{fig:stre-lamb}(b) for $\Delta t = 0$. Shortly after the transition time $t_{0}\equiv \log(N_{0})/\omega$, $\lambda(t)$ has square-root time-dependence. By Taylor-expanding Eq. \eqref{eq:lambt} about $t_{0}$ to the lowest order, one can show that $\lambda(t,\Delta t)\approx k_{\lambda}\sqrt{t-t_{0}-\Delta t}$, where $k_{\lambda}=gbN_{0}\sqrt{2\omega}/[(l_{d}-l_{m})(\langle l \rangle+d_{0})]$ \cite{SI}. This allows us to approximate $f(t|\Delta t)$ as a Gaussian
\begin{equation}
\label{eq:fGaus}
f(t|\Delta t)\approx \mathcal{N}\left[ t_{0}'+\Delta t, \sigma_{t}^{2} \right],
\end{equation}
having mean $t_{0}'+\Delta t$, with $t_{0}'=t_{0}+\Gamma(5/3)(2k_{\lambda}/3)^{-2/3}$, and variance $\sigma_{t}^{2}=(2k_{\lambda}/3 )^{-4/3}\left[\Gamma(7/3)-\Gamma^{2}(5/3) \right]$. Integrating the joint PDF $f(t|\Delta t)\mathcal{N}(0,\delta^{2}_{\Delta t})$ over $\Delta t$ yields the PDF associated with observing the first extrusion at time $t^{*}$:
\begin{equation}
  \label{eq:pt}
  p(t^{*})=\ \mathcal{N}\left[ t_{0}',\  \sigma_{t}^{2}+\delta_{\Delta t}^{2} \right].
\end{equation}
This is is displayed in Fig. \ref{fig:statistics}(b) (solid lines) and is in excellent agreement with the numerical data. Similarly, we can calculate the probability distribution associated with the extrusion occurring at position $x^{*}$. From previous considerations, one can reasonably assume the extrusion location to be uniformly distributed within the P zone. Thus the conditional PDF associated with observing the first extrusion at time $t$ and position $x$ is $f(x,t|\Delta t)=f(t|\Delta t)/L^{*}$, with $-L^{*}/2 \le x \le L^{*}/2$. Integrating over $t$ and $\Delta t$ yields:
\begin{equation}\label{eq:px}
    p(|x^{*}|)=\left( \frac{2}{3}k_{\lambda}k_{x}^{3} \right)^{\frac{1}{3}}\Gamma \left[\frac{2}{3},\frac{2}{3}k_{\lambda}k_{x}^{3}|x^{*}|^{3}\right],
\end{equation}
which again agrees well with the numerical data [Fig. \ref{fig:statistics}(a)]. Here, $k_{x}=2/(bN_{0}\sqrt{2\omega})$ and $\Gamma[\cdot,\cdot]$ is the incomplete Gamma function.  A detailed derivation of Eqs. \eqref{eq:pt} and \eqref{eq:px} can be found in Ref. \cite{SI}. Finally, Figs. \ref{fig:statistics}(d)-\ref{fig:statistics}(f) show the probability distributions of the extrusion position, extrusion time and critical stress for the original planar colonies (e.g., Fig. \ref{fig:snapshots}). Despite the mechanical interactions being more complex in planar colonies \cite{You:2018,SI}, the physical picture emerging from the simulations is nearly identical to that discussed for chainlike colonies.


In this work, we have proposed a theoretical picture for the mono- to multilayer transition in sessile bacterial colonies, using a toy model of growing spherocylindrical cells. In particular, we have focused on the interplay between stress distribution and the inherent randomness of cell division and demonstrated how this leads to a hybrid deterministic-stochastic transition, characterized by an ensemble of critical states emerging above a deterministic stress threshold. Whereas this transition originates from mechanisms similar to those investigated in Refs. \cite{Grant:2014,Beroz:2018}, some notable differences must be highlighted. In Ref. \cite{Grant:2014}, colonies are sandwiched between agar and glass and the restoring forces, competing with the growth-induced in-plane stresses, arise from the vertical compression of the agar instead of molecular adhesion. This results in the appearance of radial frictional forces that render the transition less stochastic, while introducing a sensitive dependence on the agar material properties. Conversely, Beroz {\em et al}. \cite{Beroz:2018} focus mainly, but not exclusively, on the post-transitional dynamics and illustrate how, once a cell is extruded form the first layer, can serve as a fulcrum for the rotation of neighboring cells.  This gives rise to an inverse domino effect that results in the formation of a core of vertical cells expanding from the center of the colony.

Most of our predictions are amenable to experimental scrutiny. The distributions of extrusion positions and times [Eqs. \eqref{eq:pt} and \eqref{eq:px}, and Fig. \ref{fig:statistics}] can be readily extracted from experiments on monoclonal colonies freely expanding on a Petri dish. The technology for {\em in situ} stress measurements in bacterial systems is still in its infancy \cite{Oldewurtel:2015,Chu:2018}; thus direct experimental detection of the P zone appears precluded at this stage. However, some of its properties can be indirectly inferred from the distribution of the extrusion positions and the time of the extrusion event in experiments. For instance, experiments can be designed to validate that the variance of both distributions, hence the relative size of the P zone, decreases with the growth rate.

\acknowledgments

Z. Y., D. J. G. P. and L. G. are supported by The Netherlands Organization for Scientific Research (NWO/OCW) as part of the Frontiers of Nanoscience program and the Vidi scheme. A. S. was supported by the ATTRACT Investigator Grant of the Luxembourg National Research Fund.

\end{document}


\title{Supplemental Material: Mono- to Multilayer Transition in Growing Bacterial Colonies}

\author{Zhihong You}
\affiliation{Instituut-Lorentz, Universiteit Leiden, P.O. Box 9506, 2300 RA Leiden, The Netherlands}
\author{Daniel J. G. Pearce}
\affiliation{Instituut-Lorentz, Universiteit Leiden, P.O. Box 9506, 2300 RA Leiden, The Netherlands}
\author{Anupam Sengupta}
\affiliation{Physics and Materials Science Research Unit, University of Luxembourg, 162 A, Avenue de la Faïencerie, L-1511 Luxembourg City, Luxembourg}
\author{Luca Giomi}
\email{giomi@lorentz.leidenuniv.nl}
\affiliation{Instituut-Lorentz, Universiteit Leiden, P.O. Box 9506, 2300 RA Leiden, The Netherlands}

\maketitle

\section{Hard rod model}

\begin{figure}[t]
\centering
\includegraphics[width=\columnwidth]{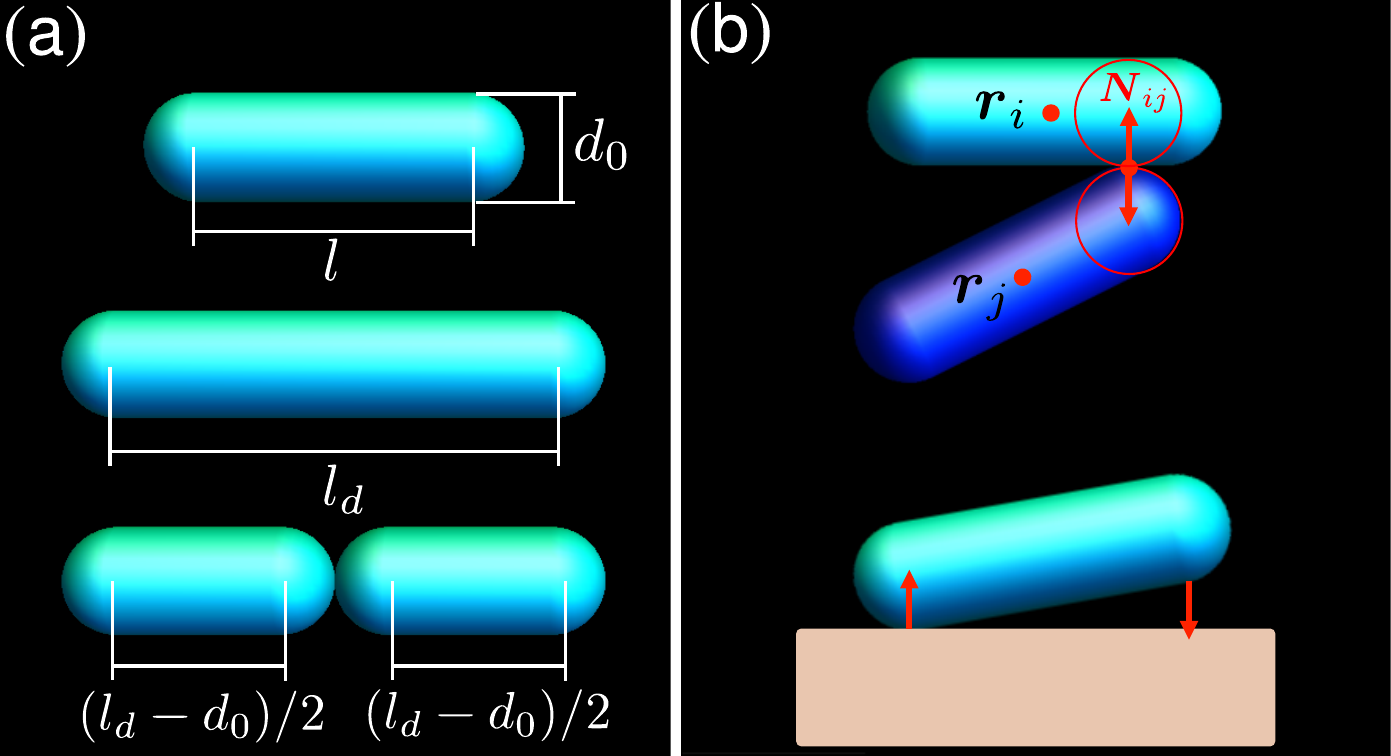}
\caption{\label{fig:md}Schematics of hard rod model. (a) Cells grow in length and, once they reach a maximal length $l_{d}$, they divide into two identical daughter cells. (b) Each cell experiences repulsive forces due to the steric interaction with the other cells (top) and the substrate (bottom), as well an attractive force from the substrate, reproducing the effect of adhesion molecules.}
\end{figure}

Each bacterium is modeled as a spherocylinder with a fixed diameter $d_{0}$ and a time-dependent length $l$ (excluding the caps on both ends, see Fig. \ref{fig:md}a), growing in a three dimensional space \cite{Farrell:2013,You:2018}. Cell growth and division are modeled as follows. The length $l_i$ of the $i$-th cell ($i=1,2,\ldots$), increases linearly in time with rate $g_{i}$ and, after having reached the division length $l_{d}$, the cell divides into two identical daughter cells (Fig. \ref{fig:md}a). In order to avoid synchronous divisions, the growth rate is randomly chosen in the interval $g/2\le g_{i} \le 3g/2$, with $g$ the average growth rate. Immediately after duplication, the daughter cells have the same orientation as the mother cell, but independent growth rates. 

The position $\bm{r}_{i}$ and the orientation $\bm{p}_{i}$ ($|\bm{p}_{i}|=1$) of $i-$th cell, are governed by the following over-damped equations:
\begin{subequations}\label{eq:rods}
\begin{align}
\frac{{\rm d}\bm{r}_{i}}{{\rm d}t} &= \frac{1}{\zeta l_i}\,\left( \sum_{j=1}^{N_{i}^{c}}\bm{F}^{c}_{ij} + \sum_{\alpha= \pm 1}\bm{F}^{s}_{i\alpha} + \boldsymbol{\eta}_{i}\right),\\
 \frac{{\rm d}\bm{p}_{i}}{{\rm d}t} &= \frac{12}{\zeta l_i^{3}} \bm{M}_{i}\times \bm{p}_{i} \\
\bm{M}_{i}&=\sum_{j=1}^{N_{i}^{c}}(\bm{r}_{ij}\times\bm{F}^{c}_{ij}) + \sum_{\alpha=\pm 1}(\bm{r}_{i\alpha}\times\bm{F}^{s}_{i\alpha}).   \end{align}
\end{subequations}
The first term on the right-hand side of Eq. (\ref{eq:rods}a) represents the steric force due to the contact between the $i-$th and $j-$th cell, with $N_{i}^{c}$ the total number of cells in contact with the $i-$th. The points of contact have positions $\bm{r}_{ij}$ with respect to the center of mass of the $i-$th cell and apply Hookean forces $\bm{F}^{c}_{ij}=k_{c} h_{ij}\bm{N}_{ij}$, where $k_{c}$ is an elastic constant, $h_{ij}$ is the overlap distance between the cells and $\bm{N}_{ij}$ their common normal unit vector (Fig. \ref{fig:md}b). The second term on the right-hand side of Eq. (\ref{eq:rods}a) is the force associated with the interaction between the cell caps, at positons $\bm{r}_{i\alpha}=\alpha l_{i}\bm{p}_{i}/2$ with respect to the center of mass, and the substrate. This force can be either repulsive or attractive, depending on the positions of the cap centroids, in such a way to model the substrate impenetrability as well as the vertical restoring force. Thus $\bm{F}_{i\alpha}^{s}= k_{s}(d_{0}/2-z_{i\alpha})\bm{\hat{z}}$, where $z_{i\alpha}$ is the $z-$coordinate of the caps, if $z_{i\alpha}<d_{0}/2$, or $\bm{F}_{i\alpha}^{s}=k_{a}l_{i}(d_{0}/2-z_{i\alpha})\bm{\hat{z}}$, if $d_{0}/2<z_{i\alpha}<d_{0}/2+r_{a}$, where $r_{a}$ is the range of the adhesive force (Fig. \ref{fig:md}b). For $z_{i\alpha}>d_{0}/2+r_{a}$ the adhesion molecules break and the cell is fully disconnected from the substrate. Here, the vertical restoring force is proportional to the cell length, under the assumption that the number of adhesion molecules is proportional to the area of the cells \cite{Beroz:2018}. The last term in Eq. (\ref{eq:rods}a) represents a weak random random force uniformly distributed in the interval $[-10^{-6},10^{-6}]\,{\rm N}$. Despite being much smaller than any other force in Eq. (\ref{eq:rods}b), this random force prevents the formation of unrealistic straight chains of cells. Finally, Eq. (\ref{eq:rods}b) represents the rotation of the cell axis in response to the torque $\bm{M}_{i}$ as described in Eq. (\ref{eq:rods}c). The constant $\zeta$ is a drag per unit length, independent of the cell orientation.

Equations. \eqref{eq:rods} have been numerically integrated using the following set of parameter values: $d_{0}=1\,\mu$m, $k_{c}=k_{s}=10$ MPa $\mu$m, $\zeta=100$ Pa h and $r_{a}=0.01\ \mu$m\cite{Farrell:2013}. The division length $l_{d}$ varies from $3\,\mu$m to $4\,\mu$m and the growth rate varies from $1\,\mu$m/h to $2\,\mu$m/h. The integration is performed with a time step $\Delta t=10^{-6}$ h.

\section{Stress measurements}

\begin{figure}[t]
\centering
\includegraphics[width=\columnwidth]{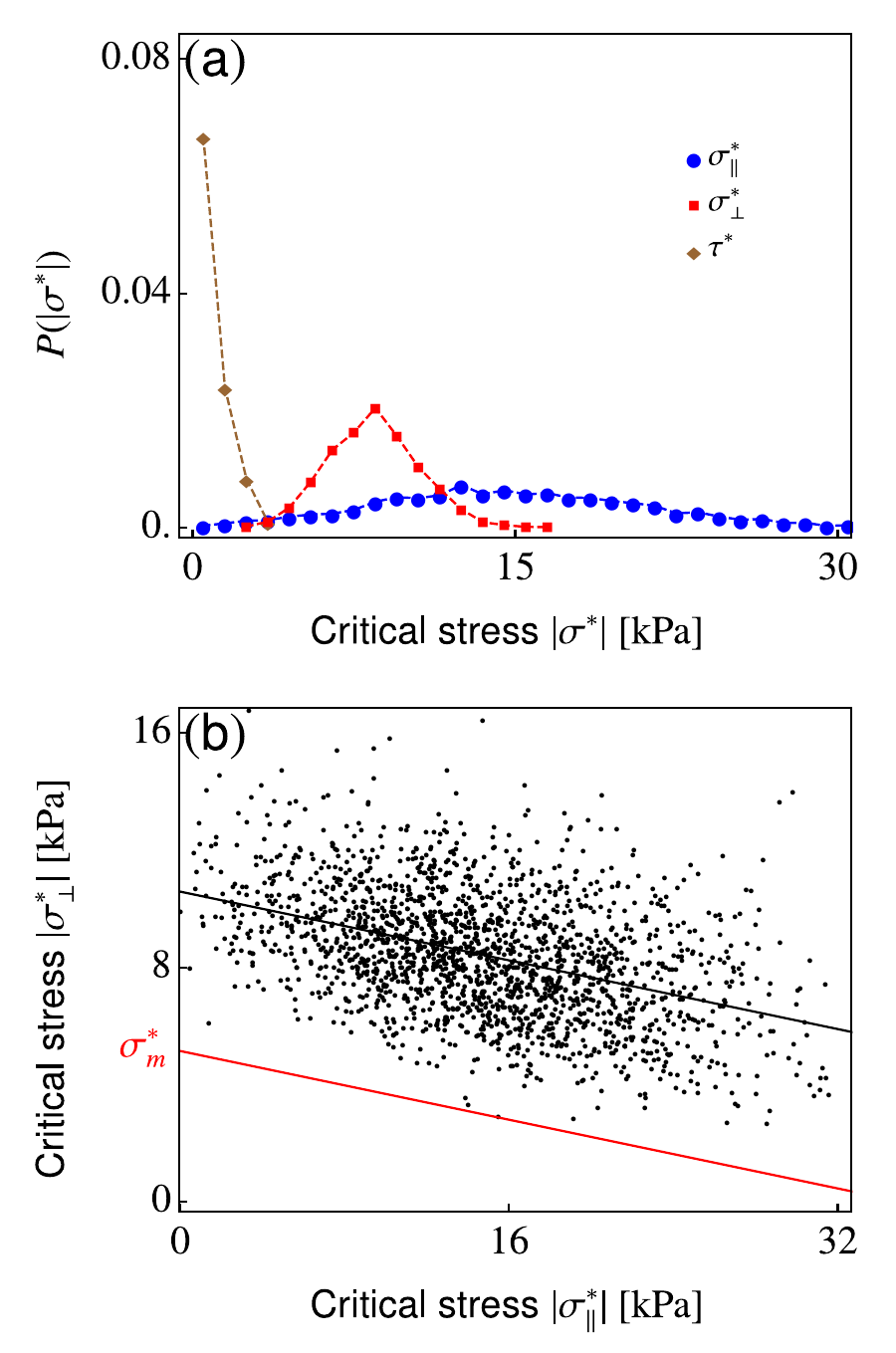}
\caption{\label{fig:stress_si}Critical stress in a planar colony at the control parameters. (a) The PDF of different components of the critical stress. (b) The two normal components of the critical stress $\sigma^{*}_{\parallel}$ and $\sigma^{*}_{\perp}$ are anti-correlated (i.e. $\sigma_{\perp}^{*} \sim -\sigma_{\parallel}^{*}$) and fall above a line of minimal critical stress $\sigma_{m}^{*}$ (in red).}
\end{figure}
The amount of stress within the colony is measured via the virial construction (see e.g. Refs. \cite{Volfson:2008,Fuentes:2013}). We focus on the in-plane stresses that trigger the mono-to-multilayer transition. Namely:
\begin{equation}
  \label{eq:virial}
  \boldsymbol{\sigma}_{i}=\frac{1}{A_{i}}\sum_{j=1}^{N_{i}^{c}}(\bm{\Pi}\cdot\bm{r}_{ij})(\bm{\Pi}\cdot\bm{F}^{c}_{ij})\;,
\end{equation}
where $A_{i}\approx d_{0}(l_{i}+d_{0})$ is the area of the $i$-th cell and $\bm{\Pi}=\bm{I}-\bm{\hat{z}}\bm{\hat{z}}$, with $\bm{I}$ the identity, is a projection operator on the $xy-$plane. In the case of a chain-like colony, such as that depicted in Fig. 2a of the main text, the $y-$components of both $\bm{r}_{ij}$ and $\bm{F}^{c}_{ij}$ vanishes and $\sigma_{xx}$ is the only nonzero component of the in-plane stress. Note that $\sigma_{xx}$ has negative values because of the extensile nature of growth-induced stress.

In the case of planar colonies, the emergence of nematic domains makes the stress anisotropic \cite{You:2018}. The in-plane stress tensor can, nevertheless, be decomposed into longitudinal and transverse components, namely:
\begin{equation}\label{eq:sigma-i}
\bm{\sigma}_{i} = 
\sigma_{i\parallel}\bm{p}_{i}^{\parallel}\bm{p}_{i}^{\parallel}+\sigma_{i\perp}\bm{p}_{i}^{\perp}\bm{p}_{i}^{\perp}+\tau_{i}(\bm{p}_{i}^{\parallel}\bm{p}_{i}^{\perp}+\bm{p}_{i}^{\perp}\bm{p}_{i}^{\parallel})\;,
\end{equation}
where $\bm{p}_{i}^{\parallel}=(p_{ix},p_{iy})$ and $\bm{p}_{i}^{\perp}=(-p_{iy},p_{ix})$. 

As shown in Fig. \ref{fig:stress_si}a, the shear component $\tau^{*}$ is always negligible, whereas both principal normal stresses affect the stability of the planar configuration. Specifically, plotting $\sigma^{*}_{\parallel}$ against $\sigma^{*}_{\perp}$, we find that the principal normal stresses are linearly related to each other (Fig. \ref{fig:stress_si}b), namely: $\sigma^{*}_{\perp}\approx -c\sigma^{*}_{\parallel}$, with $c$ a constant. In addition, all data points fall above a line with the same slope (red line in Fig. \ref{fig:stress_si}b). As a consequence, we can define an effective stress $\sigma=\sigma_{\perp}+c\sigma_{\parallel}$ for each cell, as well as a minimal effective critical stress $\sigma_{m}^{*}$ as indicated in Fig. \ref{fig:stress_si}b. Whenever $\sigma>\sigma_{m}^{*}$, a cell division can always drive the planar configuration unstable. Fig. 4f in the main text shows the probability distribution of the effective critical stress $\sigma^{*}\equiv\sigma_{\perp}^{*}+c\sigma_{\parallel}^{*}$.

\section{Stress distribution across chain-like colonies}

Eq. (2) in the main text, can be analytically derived upon taking the continuum limit. Under this assumption, the colony's mass density $\rho$, stress $\sigma_{xx}$ and velocity $v_{x}$ are related by the following equations:
\begin{subequations}\label{eq:continuum_limit}
\begin{gather}
\partial_{t}\rho + \partial_{x}(\rho v_{x}) = k_{g}\rho\;,\\
\partial_{x}\sigma_{xx}+\xi v_{x} = 0\;.	
\end{gather}	
\end{subequations}
The first equation describes the exponential increase in the cell number, with $k_{g}$ the cell division rate, whereas the second equation describes the balance between growth-induced stress and the drag force exerted by the substrate, with $\xi$ a drag coefficient. Assuming the density constant throughout the colony, yields $v_{x}=k_{g}x$. Then, solving the stress equation with free boundary conditions, e.g. $\sigma_{xx}(\pm L/2) = 0$, with $L$ the length of the colony, yields:
\begin{equation}\label{eq:sigmaxx}
\sigma_{xx} = \frac{k_{g}\xi L^{2}}{8}\,\left[1-\left(\frac{2x}{L}\right)^{2}\right]\;,
\end{equation}
consistent with Eq. (2) in the main text. In order to relate the material constant $k_{g}$ and $\xi$ in Eq. \eqref{eq:continuum_limit} with the parameters of the microscopic model defined by Eqs. \eqref{eq:rods}, one can proceed as follows. 

To calculate $\xi$, consider a cell of length $l$ and position $x$, subject to a stress difference
\begin{align*}
\Delta\sigma_{xx}
&=\sigma_{xx}\left(x+\frac{l+d_{0}}{2}\right)-\sigma_{xx}\left(x-\frac{l+d_{0}}{2}\right) \\
&\approx (l+d_{0})\,\partial_{x}\sigma_{xx}\;,
\end{align*}
where we have expanded $\sigma_{xx}$ at the linear order in $l+d_{0}$. From Eq. \eqref{eq:virial}, the force resulting from the stress differences across each cell is $\bm{F}=-d_{0}\Delta\sigma_{xx}\,\bm{\hat{x}}$. The velocity, is then given by Eq. (\ref{eq:rods}a) in the form:
\begin{align}\label{eq:vxmd}	
v_{x}
&=-\frac{d_{0}}{\zeta l}\Delta\sigma_{xx} 
\approx \frac{d_{0}(l+d_{0})}{\zeta l}\,\partial_{x}\sigma_{xx} \notag \\
&\approx \frac{d_{0}(\langle l \rangle+d_{0})}{\zeta \langle l \rangle}\,\partial_{x}\sigma_{xx}\;.
\end{align}
where in the last step we have approximated $l\approx\langle l \rangle$. Comparing Eqs. (\ref{eq:continuum_limit}b) and \eqref{eq:vxmd} yields:
\begin{equation}
\label{eq:xi}
\xi\approx\frac{\zeta \langle l \rangle}{d_{0}(\langle l \rangle+d_{0})}.
\end{equation}


Analogously, the velocity of the center of center of mass of the $i-$th cell at postion $x$ is given by:
\begin{equation}
v_{x} = \sum_{j=1}^{i} g_{j} \approx i g \approx \frac{x}{\langle l \rangle + d_{0}}\,g\;,
\end{equation}
from which we find:
\begin{equation}\label{eq:kg}
k_{g} \approx \frac{g}{\langle l \rangle + d_{0}}\;.	
\end{equation}
Replacing Eqs. \eqref{eq:xi} and \eqref{eq:kg} in Eq. \eqref{eq:sigmaxx}, we find:
\begin{equation}
\label{eq:sigxxx1}
\sigma_{xx}(x)=\frac{\zeta g\langle l \rangle L^{2}}{8d_{0}(\langle l \rangle+d_{0})^{2}}\left[ 1-\left(\frac{2x}{L}\right)^{2} \right].
\end{equation}
Since $N \approx L/(\langle l \rangle + d_{0})$, one has $a=\zeta g\langle l \rangle/(8d_{0})$ and $b\approx \langle l \rangle+ d_{0}$. A Comparison between the simulated and analytical values of $a$ and $b$ can be found in Tab. \ref{tab:ab}.
\begin{table}[htbp]
\centering 
\begin{tabular}{|c|c|c|c|c|} 
  \hline
  \multirow{2}{1.7cm}{\centering Parameters [$\mu$m],[$\mu$m/h]}& \multicolumn{2}{p{3.2cm}|}{\centering $b$ [$\mu$m]} & \multicolumn{2}{p{3.2cm}|}{\centering $a$ [kPa]} \\
  \cline{2-5} & \multicolumn{1}{c|}{Simulation} & \multicolumn{1}{c|}{Analytics} & \multicolumn{1}{c|}{Simulation} & \multicolumn{1}{c|}{Analytics} \\
\hline 
$l_{d}=3$, $g=2$ & 2.8 & 3.0 & 0.038 & 0.050 \\ \hline 
$l_{d}=4$, $g=2$ & 3.5 & 3.7 & 0.052 & 0.068 \\ \hline
$l_{d}=4$, $g=1$ & 3.5 & 3.7 & 0.026 & 0.034  \\ \hline 
\end{tabular}
\caption{Comparison between the simulated and analytical values of $a$ and $b$. The simulated values are obtained by fitting the data points shown in Fig. \ref{fig:sigm_rm}.} 
\label{tab:ab} 
\end{table}

\begin{figure}[t]
\centering
\includegraphics[width=\columnwidth]{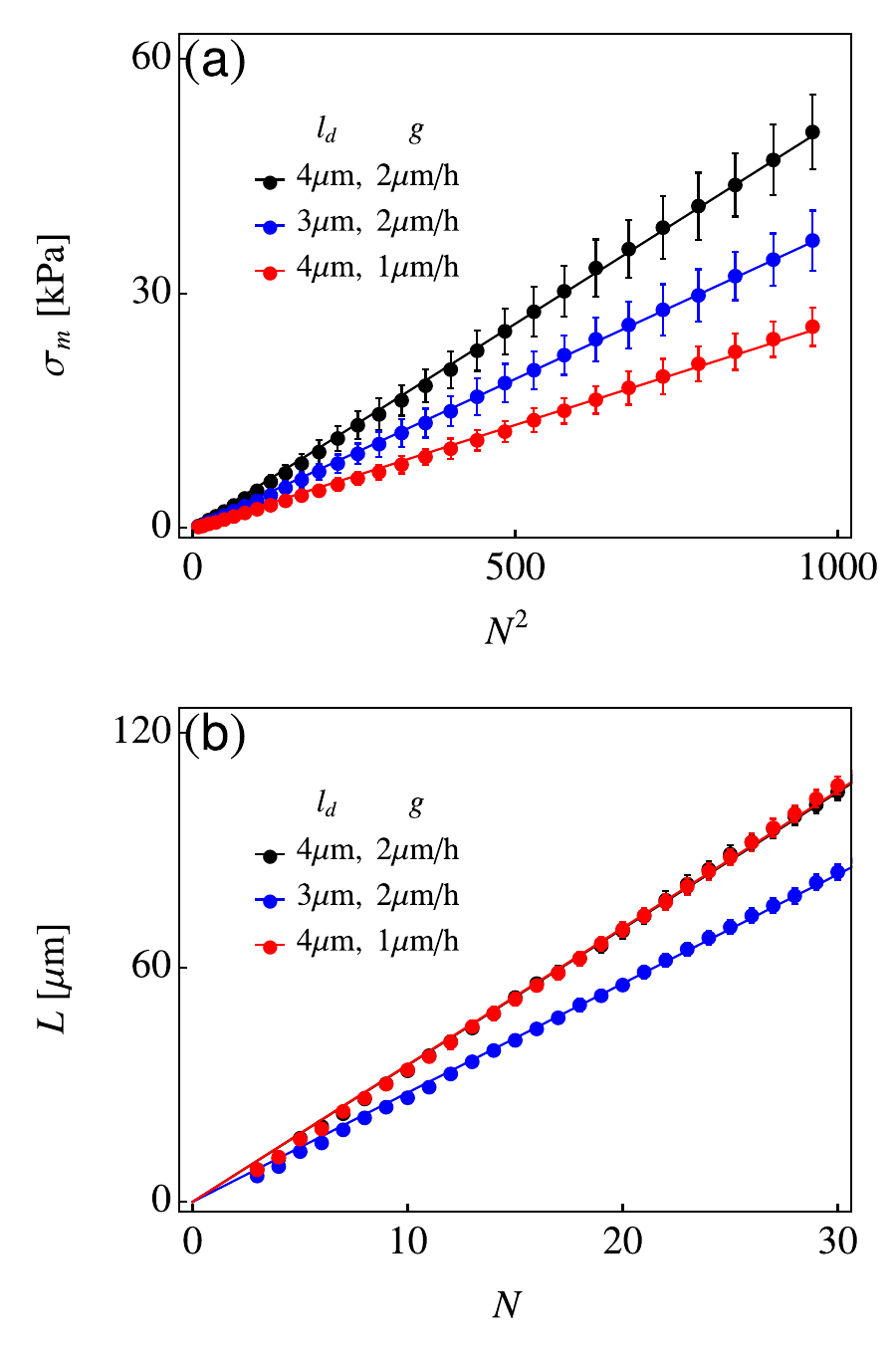}
\caption{\label{fig:sigm_rm} (a) Maximum stress $\sigma_{m}$ and (b) colony length $L$ as a function of cell number $N$, for different sets of parameters. The error bars show the standard deviation of results from $10000$ runs about the average values. Solid lines indicate the best fit, linear or parabolic, to the data points. The values of $a$ and $b$ from the fitting are shown in Tab. \ref{tab:ab}.}
\end{figure}

\section{Stochastic model of cell extrusion}

In the main text, we have explained that for $n$ cells with random lengths uniformly distributed in the interval $l_{m} \le l_{i} \le l_{d}$ and identical growth rate $g_{i}=g$, the probability of finding the first division after time $t$ is given by Eq. (3) in the main text, with rate $\lambda(n)=ng/(l_{d}-l_{m})$. In case the cell growth rate is also a random variable, that we assume uniformly distributed in the interval $g/2\le g_{i} \le 3g/2$, one can prove the rate to become
\begin{equation}
\label{eq:lambn}
\lambda(n)=\frac{1}{l_{d}-l_{m}}\,\sum_{i=1}^{n}g_{i}\;.
\end{equation}
For sufficiently large $n$ values, by the law of large numbers we have $\sum_{i=1}^{n}g_{i}\approx ng$ so that:
\begin{equation}
\label{eq:lambn1}
\lambda(n)\approx \frac{ng}{l_{d}-l_{m}}\;.
\end{equation}
Our goal is calculate the probability distribution function (PDF) associated with finding a division in the P-zone. In this case, $n$ is the total number of cells in the P-zone, which is not only time dependent, but also fluctuating due to the random cell growth. To make progress we first notice that the local stress follows a parabolic profile [Eq. (2) in the main text]:
\begin{equation}
\label{eq:sigmaxn}
\sigma_{xx}(x,N)=aN^{2}\left[ 1-\left(\frac{2x}{bN}\right)^{2} \right].
\end{equation}
The length of the P-zone is then:
\begin{equation}
\label{eq:lp}
L^{*} = \sum_{i=1}^{n}l_{i} =  b \sqrt{N^{2}-N_{0}^{2}}\;,
\end{equation}
where $N_{0}=\sqrt{\sigma_{m}^{*}/a}$ is the minimal number of cells required for the P-zone to exist. Within the P-zone, the total number of cells $n$ is a random variable, because of the random growth rate. Approximating $\sum_{i=1}^{n}(l_{i}+d_{0})\approx n(\langle l \rangle+d_{0})$, with $\langle l \rangle=(l_{d}+l_{m})/2$ the average cell length, we have $n=L^{*}/(\langle l \rangle+d_{0})$. Substituting this in Eq. \eqref{eq:lambn1} yields: 
\begin{equation}
\label{eq:lambn2}
\lambda(N) = \frac{gb}{(l_{d}-l_{m})(\langle l \rangle+d_{0})}\sqrt{N^{2}-N_{0}^{2}}\;,
\end{equation}
Notice that $\lambda$ depends on time only through the total number of cells $N$, thus $\lambda(t)=\lambda[N(t)]$. 

\begin{figure}[t]
\centering
\includegraphics[width=\columnwidth]{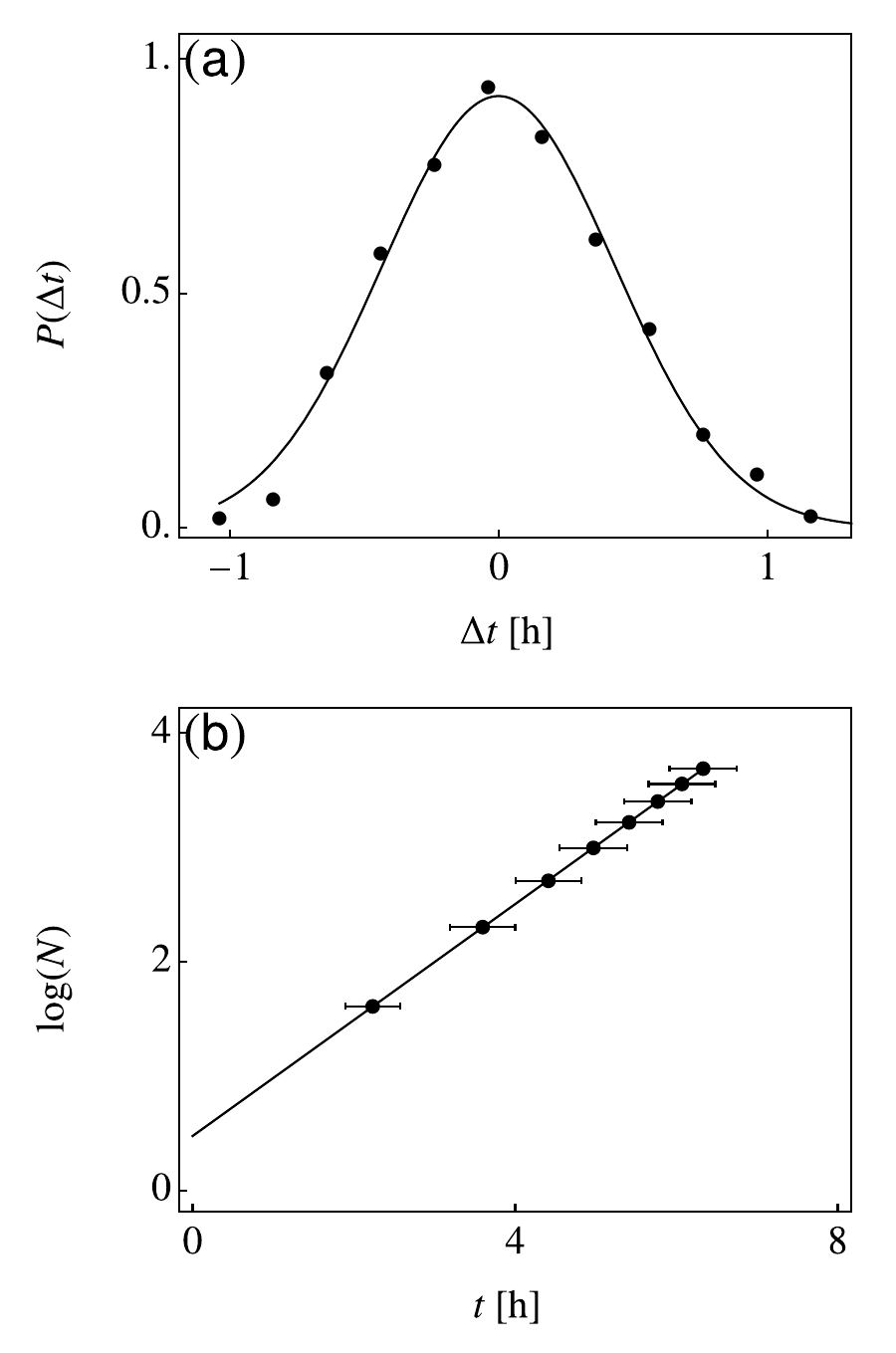}
\caption{\label{fig:n_t}Relation between cell number $N$ and time $t$ for $g=2\mu$m/h and $l_{d}=4\mu$m. (a) The time $t$ taken by the colony to attain a given population $N$ is a random variable of the form $t=\bar{t}+\Delta t$, with $\Delta t$ Gaussianly distributed (the solid line shows the best Gaussian fit). (b) $N$ increases exponentially with the average time $\bar{t}$. The black dots represent the average $\bar{t}$ and the horizontal error bars the standard deviation each associated with $10^{4}$ data points.}
\end{figure}

Because of the random growth rate, the time $t$ taken for the colony to attain a given population $N$ is a random variable of the form $t=\bar{t}+
\Delta t$. Numerically, we find that $\Delta t$ approximatively follows a Gaussian distribution $\mathcal{N}(0,\delta_{\Delta t}^{2})$ with zero mean and variance $\delta_{\Delta t}^{2}$ (Fig. \ref{fig:n_t}a). In addition, $N(\bar{t})\sim \exp(\omega \bar{t})$ (Fig. \ref{fig:n_t}b), as a consequence of the exponential growth, or equivalently $\bar{t}=\omega^{-1}\log(N)$. We can then express $t=\omega^{-1}\log(N)+\Delta t$ and $N(t,\Delta t)=\exp[\omega(t-\Delta t)]$. Replacing this in Eq. \eqref{eq:lambn2} yields:
\begin{equation}
\label{eq:lambt}
\lambda(t,\Delta t)=\frac{gb}{(l_{d}-l_{m})(\langle l \rangle+d_{0})}\sqrt{e^{2\omega(t-\Delta t)}-N_{0}^{2}}\;.
\end{equation}
As shown in Fig. 3a in the main text, at fixed $\Delta t=0$, $\lambda(t,0)\sim\sqrt{t-t_{0}}$ for $t\gtrsim t_{0}$, where $t_{0}=\omega^{-1}\log(N_{0})$ is the average time at which the P-zone first appears. Because $f(t)$ decreases exponentially with $\lambda$, most extrusions occur close to the onset. Thus, expanding $\lambda$ about $t_{0}+\Delta t$ at the lowest order yields:
\begin{equation} 
\label{eq:lambs}
\lambda(t,\Delta t)\approx k_{\lambda}\sqrt{t-t_{0}-\Delta t},
\end{equation}
where
\begin{equation}
\label{eq:klamb}
k_{\lambda}=\frac{gbN_{0}\sqrt{2\omega}}{(l_{d}-l_{m})(\langle l \rangle+d_{0})},
\end{equation}
The PDF associated with observing the first extrusion at time $t$, given its offset $\Delta t$ from the average $\bar{t}$, can then be expressed as a conditional PDF of the form:
\begin{equation}
  \label{eq:ft}
  \begin{split}
    f(t|\Delta t)=&\lambda(t,\Delta t)e^{-\int_{0}^{t}\lambda(t',\Delta t){\rm d}t'} \\
    =&k_{\lambda}(t-t_{0}-\Delta t)^{1/2}e^{-\frac{2}{3}k_{\lambda}(t-t_{0}-\Delta t)^{3/2}},
    \end{split}
\end{equation}
which is a Weibull distribution and can be approximated as a Gaussian distribution of the same mean and variance \cite{Papoulis:2002}:
\begin{equation}
\label{eq:fts}
f(t|\Delta t)\approx \mathcal{N}\left[ t_{0}'+\Delta t, \sigma_{t}^{2} \right],
\end{equation}
where $t_{0}'=t_{0}+\Gamma(5/3)(2k_{\lambda}/3)^{-2/3}$ and $\sigma_{t}^{2}=(2k_{\lambda}/3 )^{-4/3}\left[\Gamma(7/3)-\Gamma^{2}(5/3) \right]$. The joint PDF of the extrusion time $t^{*}$ and $\Delta t$ is then $f(t^{*}|\Delta t)\mathcal{N}(0,\delta_{\Delta t}^{2})$. Therefore, upon integrating over $\Delta t$:
\begin{equation}
  \label{eq:pts}
  \begin{split}
    p(t^{*})=&\int_{-\infty}^{\infty}{\rm d} \Delta t\, f(t^{*}|\Delta t)\mathcal{N}(0,\delta^{2}_{\Delta t})  \\
    =&\mathcal{N}\left[ t_{0}', \sigma_{t}^{2}+\delta_{\Delta t}^{2} \right].
    \end{split}
\end{equation}
As demonstrated by Fig. 4b in the main text, the analytical results are in good agreement with those obtained from the numerical simulations.

Since cell division occurs uniformly in the P-zone, the PDF associated with observing an extrusion at time $t$ and position $x$ is given by:
\begin{equation}
\label{eq:fxt}
f(x,t|\Delta t)
= \frac{f(t|\Delta t)}{L^{*}}
= \frac{\lambda(t,\Delta t)}{L^{*}}e^{-\int_{0}^{t}\lambda(t',\Delta t){\rm d}t'},
\end{equation}
for $-L^{*}/2<x<L^{*}/2$. From Eqs. \eqref{eq:lp} and \eqref{eq:lambn2}, $\lambda/L^{*}=g/[(l_{d}-l_{m})(\langle l \rangle+d_{0})]$. Substituting this in Eq. \eqref{eq:fxt} gives
\begin{equation}
  \label{eq:fxts}
  \begin{split}
    f(x,t|\Delta t)=&\frac{g}{(l_{d}-l_{m})(\langle l \rangle+d_{0})}\,e^{-\int_{0}^{t}\lambda(t',\Delta t){\rm d}t'} \\
    =&\frac{g}{(l_{d}-l_{m})(\langle l \rangle+d_{0})}\,e^{-\frac{2}{3}k_{\lambda}(t-t_{0}-\Delta t)^{3/2}}
    \end{split}
\end{equation}
Thus, the conditional PDF associated with observing an extrusion at position $x$, given the offset $\Delta t$, is:
\begin{equation}
  \label{eq:fx}
  \begin{split}
f(x|\Delta t)=&\int_{t_{p}(x)+\Delta t}^{\infty} {\rm d}t\,f(x,t|\Delta t) \\
  =&\int_{t_{p}(x)}^{\infty} {\rm d}t'\,f(x,t'|0) \\
  =& f(x|0),
\end{split}
\end{equation}
where we have used the transformation $t'=t-\Delta t$ and by virtue of Eq. \eqref{eq:lambt}, $f(t,|\Delta t)$ depends on $t$ and $\Delta t$ only via the combination $t-\Delta t$. Here, $t_{p}(x)$ is the average time at which position P-zone first reaches the distance $x$ from the center of the colony. From Eq. \eqref{eq:fx}, $f(x|\Delta t)$ is independent of $\Delta t$. The PDF associated with observing the first extrusion at position $|x^{*}|$ is then:
\begin{equation}
  \label{eq:pxs0}
  \begin{split}
p(|x^{*}|)&= 2\int_{t_{p}(x^{*})}^{\infty} {\rm d}t\,f(x,t|0) \\
&=\frac{2g}{(l_{d}-l_{m})(\langle l \rangle+d_{0})}\int_{t_{p}(x^{*})}^{\infty} {\rm d}t\,e^{-\frac{2}{3}k_{\lambda}(t-t_{0})^{3/2}} \\[5pt]
&=\frac{2g(2/3)^{1/3}k_{\lambda}^{-2/3}}{(l_{d}-l_{m})(\langle l \rangle+d_{0})}\Gamma \left[\frac{2}{3},\frac{2}{3}k_{\lambda}\left(t_{p}(x^{*})-t_{0}\right)^{\frac{3}{2}}\right],
  \end{split}
\end{equation}
where $\Gamma[s,x]$ is the incomplete Gamma function. It can be demonstrated that $[t_{p}(x)-t_{0}]^{1/2}=k_{x}|x|$, where $k_{x}=2/(bN_{0}\sqrt{2\omega})$. Substitute this in Eq. \eqref{eq:pxs0}, we have
\begin{equation}
  \label{eq:pxs}
  \begin{split}
    p(|x^{*}|)&=\frac{2g(2/3)^{1/3}k_{\lambda}^{-2/3}}{(l_{d}-l_{m})(\langle l \rangle+d_{0})}\Gamma \left[\frac{2}{3},\frac{2}{3}k_{\lambda}k_{x}^{3}|x^{*}|^{3}\right] \\
    &=\left( \frac{2}{3}k_{\lambda}k_{x}^{3} \right)^{\frac{1}{3}}\Gamma \left[\frac{2}{3},\frac{2}{3}k_{\lambda}k_{x}^{3}|x^{*}|^{3}\right],
  \end{split}
\end{equation}
again in good agreement with the simulations (Fig. 4a).

\section{The role of the material parameters}

\begin{figure*}[t]
\centering
\includegraphics[width=2\columnwidth]{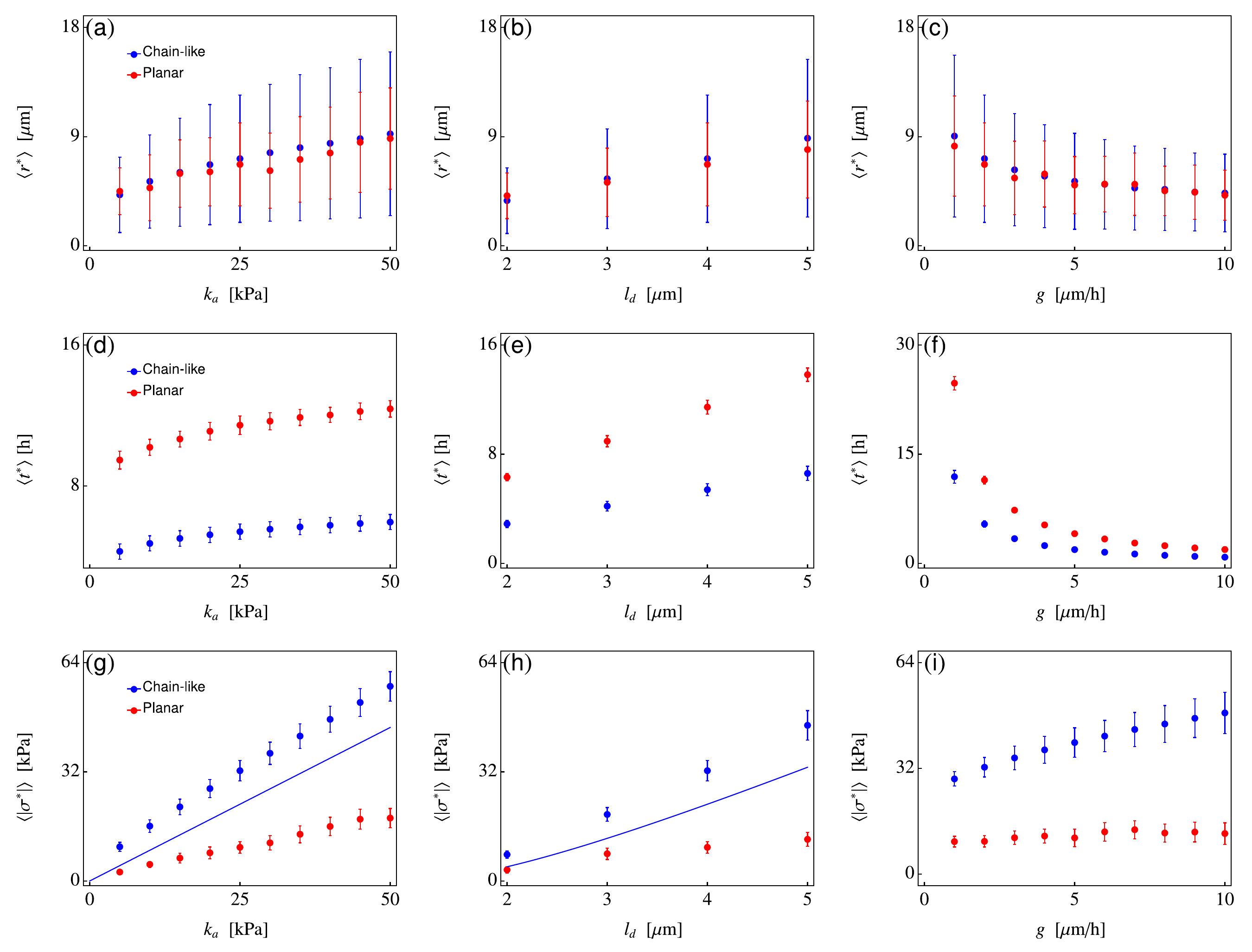}
\caption{\label{fig:trend} The average critical quantities $\langle r^{*} \rangle$, $\langle t^{*} \rangle$ and $\langle |\sigma^{*}| \rangle$ versus the material parameter $k_{a}$, $l_{d}$ and $g$ for both chain-like (blue dots) and planar (red dots) colonies. In panels (g) and (h) the blue lines indicated the values of $\sigma_{m}^{*}=d_{0}^{-1}k_{a}l_{m}^{2}/(1+l_{m}/d_{0})$, which set the lower bound of $\sigma^{*}$. The error bars show the standard deviations of $10^{4}$ ($100$) instances about the mean values for chain-like (planar) colonies.}
\end{figure*}

Fig. 4 in the main text shows the PDFs of the critical quantities $|x^{*}|$, $t^{*}$ and $|\sigma^{*}|$ for four set of material parameters. In Fig.  \ref{fig:trend} we complement this information with plots of the averages $\langle |x^{*}| \rangle$, $\langle t^{*} \rangle$ and $\langle |\sigma^{*}| \rangle$ versus the material parameter $k_{a}$, $l_{d}$ and $g$ for both chain-like (blue dots) and planar (red dots) colonies. 
In Figs. \ref{fig:trend}g and h, we compare the average critical stress $\sigma^{*}$, with the minimal critical stress, given by: $\sigma_{m}^{*}=d_{0}^{-1}k_{a}l_{m}^{2}/(1+l_{m}/d_{0})$. As expected, these quantities have the same dependence on $k_{a}$ and $l_{d}$. 

\section{Colonies with Gaussianly distributed growth rate}

\begin{figure*}[t]
\centering
\includegraphics[width=2\columnwidth]{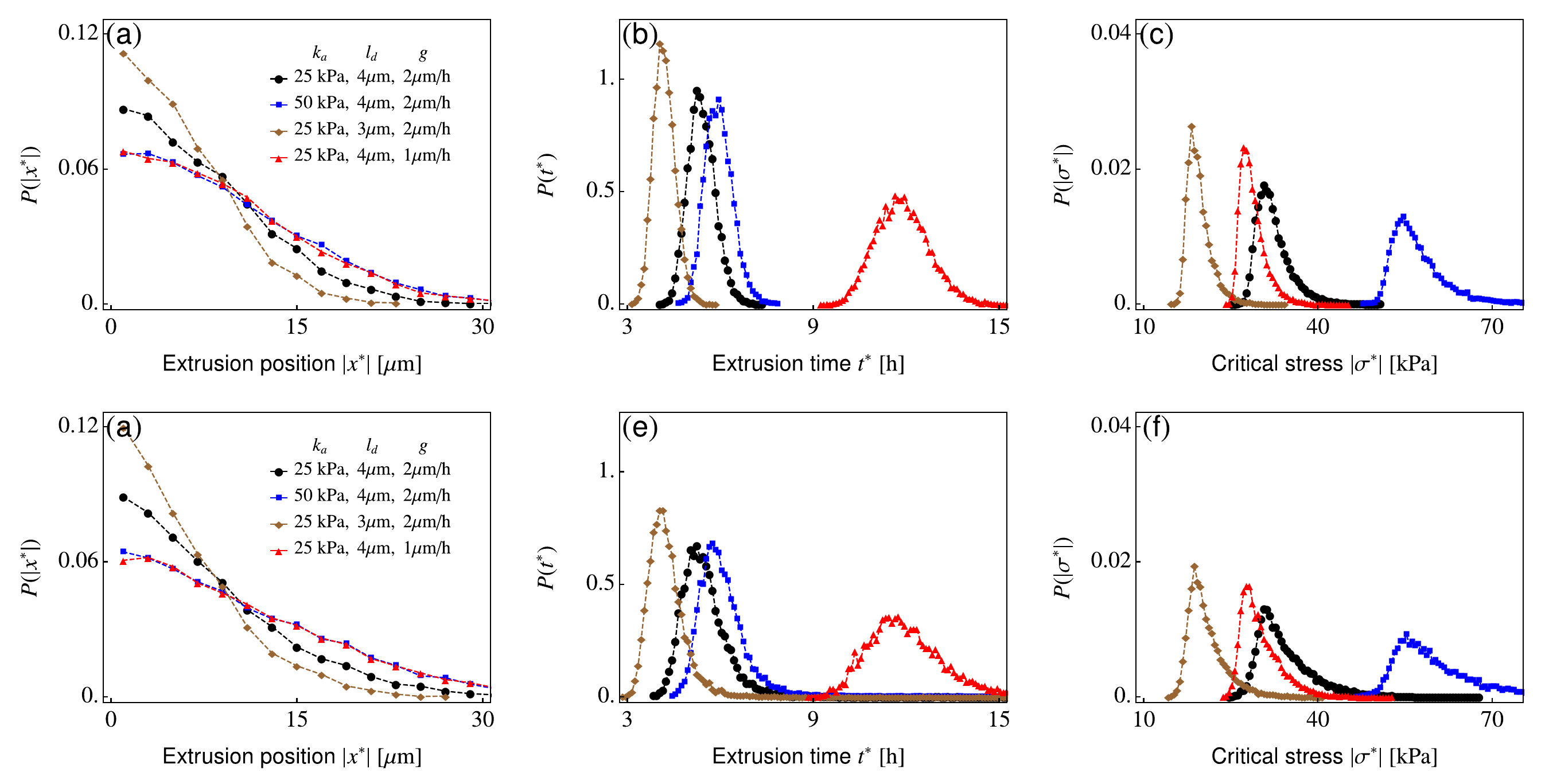}
\caption{\label{fig:stat-gaus} The same PDFs shown in Fig. 4 of the main text for uniformly (a-c) and Gaussianly (d-f) distributed growth rate, i.e. $g/2<g_{i}<3g/2$ and $g=4\mu$m/h. Both distributions have the same mean and variance. In each set of simulations, one parameter is changed compared to the control set, whose parameter values are indicated in the legends. For each set of parameters, we performed $10^{4}$ runs.}
\end{figure*}

In our toy model, the cell growth rates $g_{i}$ are drawn uniformly in the interval $g/2 \le g_{i} \le 3g/2$. In order to test the robustness of our predictions with respect to this choice, we have repeated the analysis presented in the main text using Gaussianly distributed growth rates. Fig. \ref{fig:stat-gaus} illustrate a comparison between the PDFs obtained in both cases. These are essentially indistinguishable as it is their dependence on the control parameters $k_{a}$, $l_{d}$, and $g$.
